\documentclass[floats,floatfix,showpacs,amssymb,prl,twocolumn,superscriptaddress,nofootinbib]{revtex4-2}
\usepackage{amsmath}
\usepackage{amssymb}
\usepackage{bm}
\usepackage{graphicx}
\usepackage[colorlinks=true,citecolor=blue,linkcolor=blue,urlcolor=blue]{hyperref}
\newcommand{\Om}{\ensuremath{\Omega_{\rm m}}}
\newcommand{\sig}{\ensuremath{\sigma_8}}
\newcommand{\halpha}{\ensuremath{H\alpha}}
\newcommand{\hbeta}{\ensuremath{H\beta}}
\newcommand{\oiii}{\mbox{[O\,\textsc{iii}]}}
\newcommand{\oii}{\mbox{[O\,\textsc{ii}]}}
\newcommand{\cii}{\mbox{[C\,\textsc{ii}]}}

\begin{document}
\title{Interlopers as Signal in Line Intensity Mapping}
\author{Anirban Roy}
\affiliation{Department of Physics, New York University, 726 Broadway, New York, NY, 10003, USA}
\affiliation{Center for Computational Astrophysics, Flatiron Institute, New York, NY 10010, USA}
\date{\today}

\begin{abstract}
Line intensity mapping measures the combined emission of atomic and molecular
lines from unresolved galaxies, offering a way to map cosmic structure across
enormous volumes. A single observed frequency, however, contains emission from
several spectral lines at different redshifts. These interlopers are usually
removed before cosmological inference. We show that this removal is not always
optimal. Once the possible emitting lines and their redshifts are included in
the model, interlopers become additional tracers of large-scale structure at
other epochs, projected into the coordinate system of the target line. We
formulate interloper treatment as a Fisher decision problem, requiring both
improved precision and a parameter bias below a chosen tolerance. In SPHEREx
forecasts for \halpha{} with \oiii{}, \hbeta{}, and \oii{} interlopers, and in
FYST forecasts for \cii{} with CO interlopers, $\Omega_mh^2$ and baryon
acoustic oscillation (BAO) distance measurements are comparatively robust to
astrophysical calibration errors, whereas the clustering amplitude $\sigma_8$
and the dark energy parameters $w_0$ and $w_a$ are more sensitive to the adopted
line model and priors. The baryon acoustic oscillation distance measurement
gives a simple physical picture: modeling interlopers turns two observed bands
into a transverse BAO distance ladder over $0.7\lesssim z\lesssim5.6$.
Interlopers are therefore not only contaminants to remove, but a calibrated
component of the cosmological signal model for future LIM surveys.
\end{abstract}

\maketitle
Line intensity mapping (LIM) observes the integrated emission from unresolved
galaxies over a broad redshift range, and therefore offers three-dimensional
access to large-scale structure over enormous cosmic volumes
\cite{Kovetz2017Status, Fonseca2016Lines}. A single observed frequency, however,
generally receives emission from several rest-frame lines redshifted from
different epochs of the Universe. These ``interlopers'' are normally treated as
contaminants to be masked, deprojected, nulled, or externally subtracted before
cosmological inference from the target line for which an experiment is designed
\cite{Gong2020Interlopers,Bernal2024Cleaning,Karoumpis2024CO, Roy2024-lim-cross}. This is a safe
instinct if the goal is to obtain a clean map of the target epoch and probe its
astrophysical properties along with the cosmological parameters. However, it is not automatically optimal if the goal is
to constrain cosmology tomographically.
The reason is that interloper lines are not arbitrary foregrounds analogous to
random noise or instrumental systematics. Rather, each interloper is itself a
line intensity map of large-scale structure at a different redshift, geometrically
warped into the coordinate system of the target line. If the astrophysical
modeling of the line emission is sufficiently constrained, the interloper carries
cosmological information in addition to the target line. Removing it can
therefore discard both the additional information carried by the interlopers and
the target modes sacrificed by the cleaning operation. 

This Letter formulates the resulting choice as a decision problem: should
an analysis clean the interlopers, keep and jointly model them, partially clean
and model the residuals, use external tracers, or add coeval auto- and
cross-spectra? The central point is that interloper treatment is not a fixed
preprocessing step, but an estimator choice whose optimal form depends on the
cosmological parameter, the cleaning cost, and the accuracy of the astrophysical model.
In a $\Lambda$CDM universe, cosmological parameters such as the present day
expansion rate, $H_0=100h\,{\rm km\,s^{-1}\,Mpc^{-1}}$, the physical matter
density $\Omega_mh^2$, and the clustering amplitude $\sigma_8$ determine the
redshift evolution of distances, expansion, and growth of structure \cite{Planck-cosmo-main2018}. Extensions
with dynamical dark energy introduce additional parameters, such as $w_0$ and
$w_a$ \cite{CPL-main, Linder2003-cpl}. The astrophysical processes that connect galaxy formation to line
emission are intricate and remain uncertain, whereas the impact of cosmological
parameters on large-scale structure is tightly constrained by measurements of
the cosmic microwave background and galaxy surveys \cite{Planck-cosmo-main2018, DESI-JCAP-2024}. In this Letter, we explore
the possibility of using interlopers in LIM analyses to constrain these
cosmological parameters under different astrophysical priors. A careful joint analysis of interlopers and the target signal can also probe BAO
features at multiple redshifts, extending the distance ladder to epochs that are
difficult to reach with conventional galaxy surveys. Thus, interloper-aware modeling can turn a source of contamination into additional cosmological leverage for LIM surveys.
\begin{figure*}[t]
  \centering  \includegraphics[width=\textwidth]{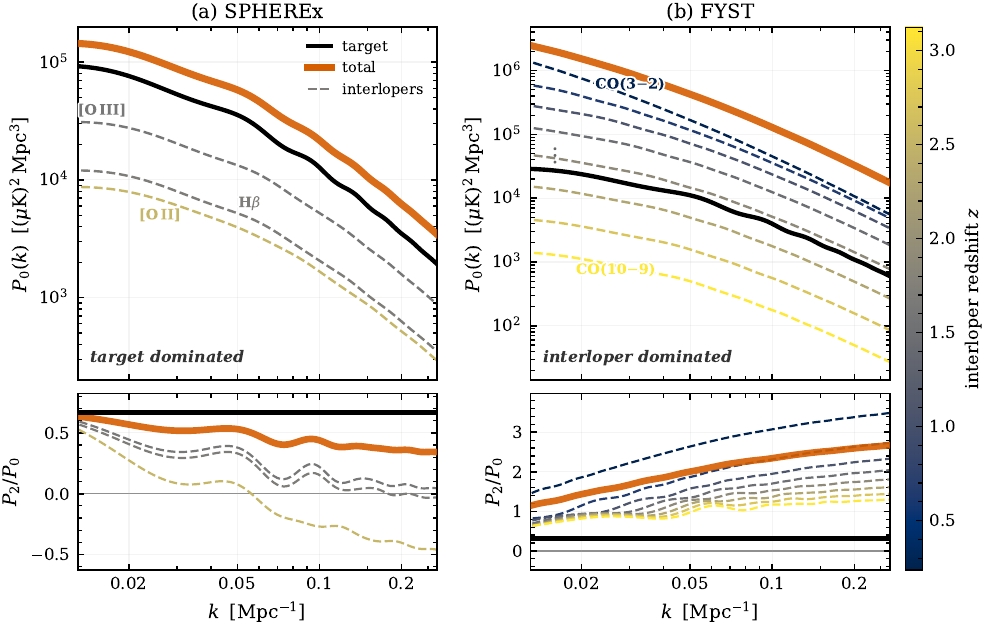}
   \caption{
    Projected monopole $P_0(k)$ (top), and quadrupole ratio $P_2/P_0$ (bottom),
    on a shared $k$ axis. The left panel shows a target-dominated survey,
    SPHEREx-like H$\alpha$ at $z=1$; the right panel an interloper-dominated
    one, FYST-like \cii{} at $z=5.8$. Black curves show the target alone,
    dashed curves the individual interlopers colored by emission redshift, and
    the thick vermillion curve the observed total, summed \emph{after} the
    Alcock--Paczy\'nski projection of Eq.~(\ref{eq:pobs}). On the left the
    target dominates and $P_2/P_0$ stays near its Kaiser value; on the right
    the CO ladder outshines the faint \cii{} target and pulls $P_2/P_0$ far
    from it. This projected anisotropy is the interloper signal that carries
    the distance information.%
  }
  \label{fig:mechanism_grid}
\end{figure*}

\emph{A decision criterion.---}
A joint analysis that keeps interlopers is useful only if their added
cosmological information outweighs the cost of modeling them. For a cosmological parameter $p$, cleaning removes non-target emission, but the
same operation can also reduce the usable target information by masking voxels,
projecting out modes, reducing the effective volume, or increasing the
covariance of the cleaned map \cite{Gong2020Interlopers, Bernal2024Cleaning, Karoumpis2024CO}. Joint modeling retains the full map at the price of
astrophysical nuisance parameters such as line amplitudes, clustering biases,
and shot noise across several redshifts. We quantify this balance with the Fisher information, taken here
as the inverse marginalized variance on $p$ after these nuisance parameters are
included in our analysis. The scalar $F$ below denotes the final information in the parameter
direction $p$, and the forecasts use the full multi-parameter Fisher matrix.
Let $F_{\rm target}$ be the information available in an ideal target map. A
cleaned analysis yields
\begin{equation}
F_{\rm clean}=F_{\rm target}-F_{\rm lost},
\end{equation}
where $F_{\rm lost}\ge0$ is the target information sacrificed by cleaning. If
the interlopers are instead kept and modeled, the same observed map contains
additional large-scale structure at other redshifts. After marginalizing over
the nuisance parameters, we can write
\begin{equation}
F_{\rm joint}=F_{\rm target}+F_{\rm interloper}^{\rm eff}.
\end{equation}
The effective interloper term $F_{\rm interloper}^{\rm eff}$ need not be
positive for every parameter, since extra lines can also introduce degeneracies
with astrophysics. Comparing the two analyses, and before accounting for
systematic bias, joint modeling is favored when
\begin{equation}
F_{\rm interloper}^{\rm eff}+F_{\rm lost}>0 .
\end{equation}
In this case, precision alone does not make an estimator useful, because a miscentered astrophysical
prior can shift the recovered cosmology even as the statistical error shrinks.
We therefore impose the stronger requirement that an estimator deliver both a
precision gain and a bounded parameter bias,
\begin{equation}
G_p\equiv \frac{\sigma_{\rm clean}(p)}{\sigma_{\rm alt}(p)}>1,
\qquad
\frac{|\Delta p|}{\sigma_{\rm alt}(p)}<B_{\rm max},
\label{eq:decision}
\end{equation}
where ``alt'' denotes the alternative analysis, such as joint modeling, a hybrid
of cleaning and residual modeling, cleaning by using external tracers, or an auto- plus
cross-spectrum combination. As a reference, we adopt $B_{\rm max}=0.5$, a half-$\sigma$
tolerance that keeps parameter shifts subdominant to the statistical error; the
conclusions are unchanged for the stricter choice $B_{\rm max}=0.25$. Thus
there is no single best interloper strategy for all parameters. Cleaning is
preferred when interlopers give little precision gain or produce too much bias,
whereas modeling is preferred when they reduce the error bar and remain within
the bias tolerance.\\
\emph{Modeling the signal.---}
Following the standard treatment of interloper projection in LIM
\cite{LidzTaylor2016Interlopers,Gong2020Interlopers}, with the Alcock-Paczynski
Jacobian \cite{Ballinger1996AP,SeoEisenstein2007BAO} and linear Kaiser
redshift-space distortions \cite{Kaiser1987RSD}, we model the auto power
spectrum as
\begin{align}
P_{\rm obs}(k,\mu)
&=P_N+\sum_L
\frac{1}{q_{\perp,L}^2q_{\parallel,L}}
\Bigl[
I_L^2
\left(b_L+f_L\mu_L^2\right)^2
\nonumber\\
&\hspace{2.2cm}\times
P_{\rm m}(k_L,z_L)
+P_{{\rm shot},L}
\Bigr],
\label{eq:pobs}
\end{align}
where $P_N$ is the instrumental noise power spectrum and the index $L$ labels the spectral line contributing to the observed band,
including the target and all interlopers.  The factors
$(q_{\perp,L},q_{\parallel,L})$ are the transverse and radial projection
factors from the line redshift to the target coordinate system.  Thus
Eq.~(\ref{eq:pobs}) is a projected sum, not a same-redshift simple sum:
each line is evaluated at
$k_{\perp,L}=k_\perp/q_{\perp,L}$ and
$k_{\parallel,L}=k_\parallel/q_{\parallel,L}$ before the line powers are
added.  Here
$k_L=(k_{\perp,L}^2+k_{\parallel,L}^2)^{1/2}$,
$\mu_L=k_{\parallel,L}/k_L$, and
$f_L\equiv f(z_L)=d\ln D/d\ln a$ is the linear growth rate at the
redshift of line $L$. The matter power spectrum $P_{\rm m}(k_L,z_L)$ is computed with the Boltzmann code CLASS \cite{class}. We constrain four cosmological parameters
$(\Om h^2,\sig,w_0, w_a)$ using SPHEREx- and FYST-like LIM surveys \cite{FYST, Spherex-Bock-2025}. For each line we include astrophysical nuisance parameters
$(\ln I_L,\ln b_L,\ln P_{{\rm shot},L})$, where $I_L$ is the mean line intensity,
$b_L$ is the luminosity-weighted linear bias, and $P_{{\rm shot},L}$ is the
Poisson shot-noise power.  These are global calibration factors shared by each rest-frame line across the redshift bins; the forecast therefore assumes that the chosen luminosity-SFR
family supplies the redshift evolution. The Fisher matrix for the cosmological and astrophysical parameters is \cite{Tegmark1997,SeoEisenstein2003, AmaraRefregier2008}
\begin{equation}
F_{\alpha\beta}=\sum_{z,k,\mu}
\frac{V_z k^2\Delta k\Delta\mu}{4\pi^2}
\frac{\partial_\alpha P_{\rm obs}\,\partial_\beta P_{\rm obs}}
{P_{\rm obs}^2}+\Pi_{\alpha\beta},
\label{eq:fisher}
\end{equation}

\begin{figure}
  \centering
\includegraphics[width=\columnwidth]{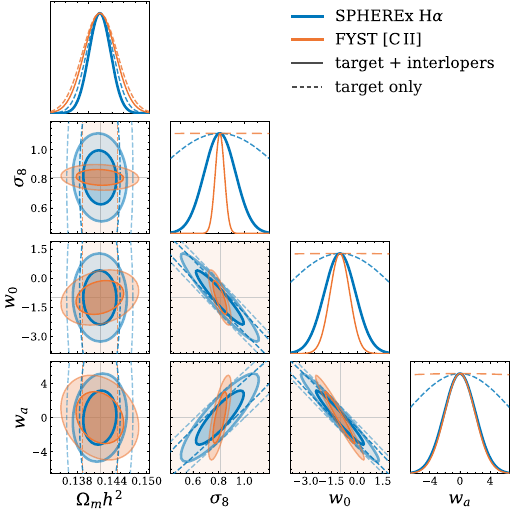}
\caption{
Marginalized $68\%$ and $95\%$ constraints for one SPHEREx \halpha{} band
(blue) and one FYST \cii{} band (orange). Filled contours jointly model targets
and interlopers; dashed contours use the target alone, with interlopers
contributing only to the covariance. Line amplitudes carry $5\%$ priors and are
marginalized. Joint modeling uses the interloper power spectra as additional
redshift tracers, improving constraints and changing the directions of degeneracy,
most visibly for the faint \cii{} target.}
  \label{fig:triangle}
  \label{fig:prior_accuracy}
\end{figure}
with $\partial_\alpha\equiv\partial/\partial\theta_\alpha$. Here the sum runs over redshift bins and Fourier bins. The indices $\alpha$ and
$\beta$ label parameters in the joint cosmological and astrophysical parameter
vector, $V_z$ is the comoving volume of each redshift bin, and
$\Pi_{\alpha\beta}$ adds Gaussian priors on nuisance parameters. If the fiducial nuisance model leaves a residual
$\Delta P_{\rm obs}\equiv P_{\rm true}-P_{\rm model}$, the induced parameter
shift is computed with the standard linearized Fisher-bias expression
\cite{Knox1998Reionization,HutererTakada2005,AmaraRefregier2008},
\begin{equation}
\Delta\theta_\alpha=(F^{-1})_{\alpha\beta}
\sum_{z,k,\mu}
\frac{V_z k^2\Delta k\Delta\mu}{4\pi^2}
\frac{\partial_\beta P_{\rm obs}\Delta P_{\rm obs}}{P_{\rm obs}^2}.
\label{eq:bias}
\end{equation}
The line moments entering $I_L$, $b_L$, and $P_{{\rm shot},L}$ are
generated from halo-mass-function moments and SFR-to-line-luminosity relations \cite{Roy-limpy-2023},
as detailed in the End Matter. The SPHEREx case uses the empirical optical
line model and the FYST case uses empirical \cii{} and CO luminosity scalings.
The main numerical example is a SPHEREx-like \halpha{} survey over
$0.7<z<1.6$ with \oiii{}, \hbeta{}, and \oii{} interlopers with
$R\simeq 40$, a $6.2$ arcsec beam scale, and $k_{\max}=0.25\,{\rm Mpc}^{-1}$ \cite{Gong2016OpticalLines,Crill2024SPHEREx}.
Figure~\ref{fig:mechanism_grid} shows the projected power spectrum
that allows interlopers to carry cosmological information. It contrasts a target-dominated \halpha{} map at $z=1$,
representative of SPHEREx, with an interloper-dominated \cii{} map at
$z=5.8$ for EoR-Spec at FYST, using the projected monopole $P_0(k)$ and quadrupole
ratio $P_2/P_0(k)$. Each interloper is evaluated at the wavevector set by its own redshift and then projected into the target coordinate system, producing the anisotropic mapping in Eq.~(\ref{eq:pobs}). When the target dominates, the observed quadrupole remains close to the target Kaiser prediction. When the interlopers dominate, as in the FYST \cii{} case with a bright CO ladder, the projected total acquires a distinct quadrupole shape. This anisotropy is the observable imprint that allows interlopers to carry distance information rather than behaving as featureless contamination.
\begin{figure}[t]
  \centering
  \includegraphics[width=\columnwidth]{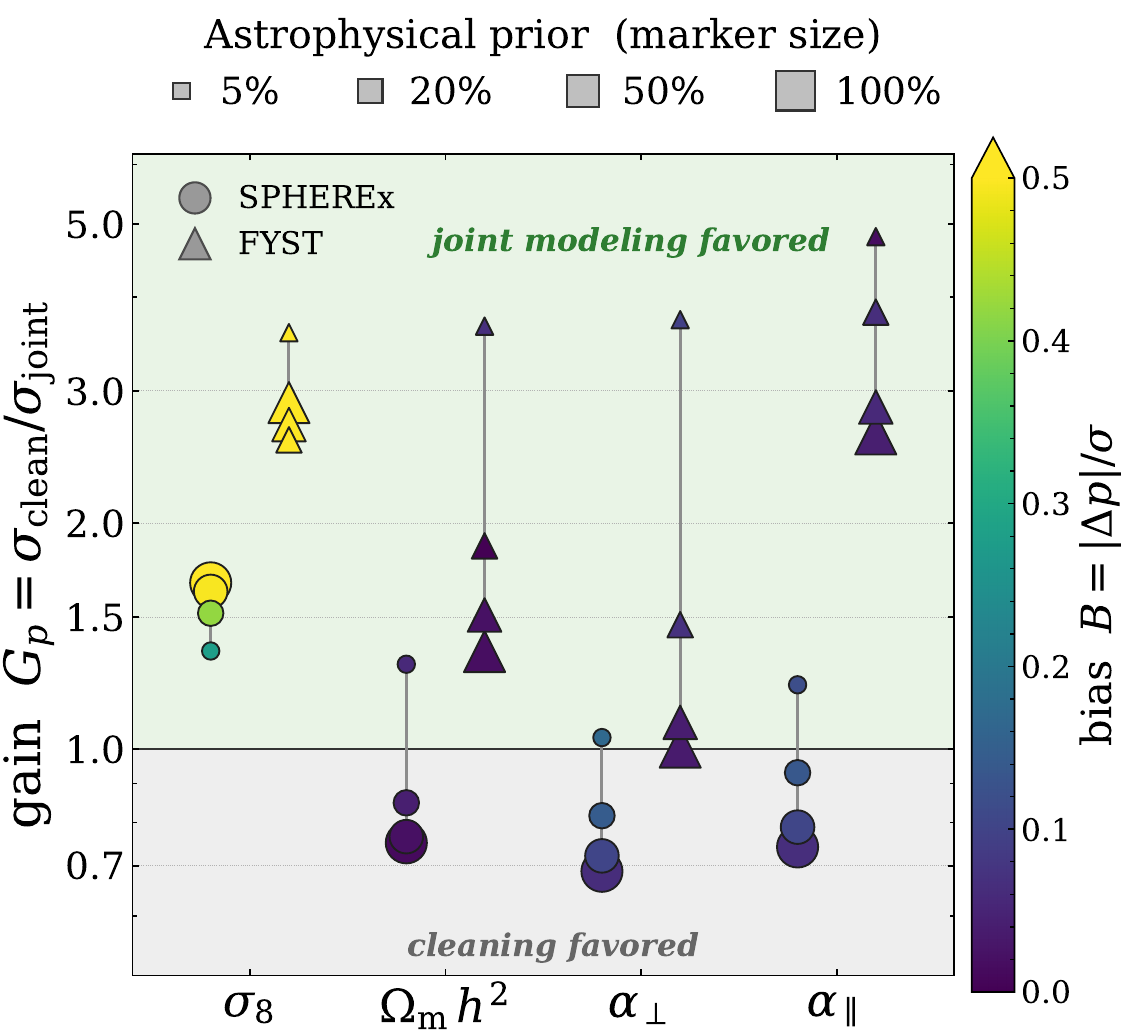}
  \caption{Precision gain $G_p$ for modeling interlopers instead of using a cleaned
target-only analysis, shown for SPHEREx \halpha{} (circles) and FYST
\cii{}$+$CO (triangles). Marker size gives the external astrophysical prior
width, from $5\%$ to $100\%$; color gives the induced bias
$B=|\Delta p|/\sigma$, capped at $B_{\rm max}=0.5$. Points above $G_p=1$ favor
joint modeling, while points below favor cleaning. Modeling often improves
precision, while $\sigma_8$ is bias limited in this test.
}
  \label{fig:scorecard}
  \label{fig:constraint_compare}
\end{figure}

\emph{Cosmology from a single band.---}
We first show that one observed LIM band can contain cosmological information
from several redshifts. In
Figure~\ref{fig:triangle} we show the forecasts for cosmological parameters from a single
SPHEREx \halpha{} band and a single FYST \cii{} band, each split into five
redshift bins that share the cosmological parameters. Modeling the target
\emph{and} its interlopers jointly (filled contours) is compared with fitting
the target alone while the interlopers remain present in the data (dashed).
Treating the interlopers as extra tracers of the same cosmology at other
redshifts, projected into one observed band, shrinks
$\sigma(\sig)$ by a factor of ${\approx} 4$ for the bright \halpha{} target and
by a factor of ${\approx}190$ for the faint \cii{} target, and tightens $w_0$
and $w_a$ by up to a factor of ${\approx}50$. These factors are relative to the unmodeled case and Fig.~\ref{fig:scorecard}
uses the cleaned baseline. Keeping the interlopers also
changes the parameter degeneracies. In the FYST case, the faint \cii{} target
alone provides weak leverage on the evolution of dark energy, while the CO
interlopers add lower redshift structure that helps constrain $w_0$ and $w_a$.
The gain is therefore largest exactly where the target is weakest, and the
interlopers become the dominant source of information rather than a contaminant.

\emph{Astrophysical modeling uncertainties.---} The gain from keeping interlopers is only useful if astrophysical errors do not significantly bias the inferred cosmology. Figure~\ref{fig:scorecard} applies the decision criterion of
Eq.~(\ref{eq:decision}) to the SPHEREx flagship case and to a FYST
\cii{}$+$CO case. For each parameter, we show the precision gain $G_p$ from
modeling the interlopers instead of using a cleaned target-only analysis. Marker
size gives the assumed external prior on the line astrophysics, from $5\%$ to
$100\%$. Marker color shows the resulting parameter bias, in units of the final
statistical error, when the true line amplitude is offset from the assumed
calibration by half of that prior width. Modeling gives smaller statistical errors ($G_p>1$) over most of the prior
range, with larger gains when the astrophysical priors are tighter. The figure highlights two distinct roles of the astrophysical prior. Its width
sets how much the line model can adjust to the data and therefore controls the
statistical error, while its accuracy controls whether the recovered cosmology is
biased. In this example, $\Omega_m h^2$ and the parameters of BAO dilation 
$(\alpha_\perp,\alpha_\parallel)$ are bias safe for all priors, but
$\sigma_8$ is improved in precision but limited by the bias due to astrophysical uncertainty. A tight but miscentered prior can, however, increase the bias on cosmological
parameters, because it restricts the likelihood from self-calibrating the interloper amplitude.

\begin{figure}
  \centering
  \includegraphics[width=0.9\columnwidth]{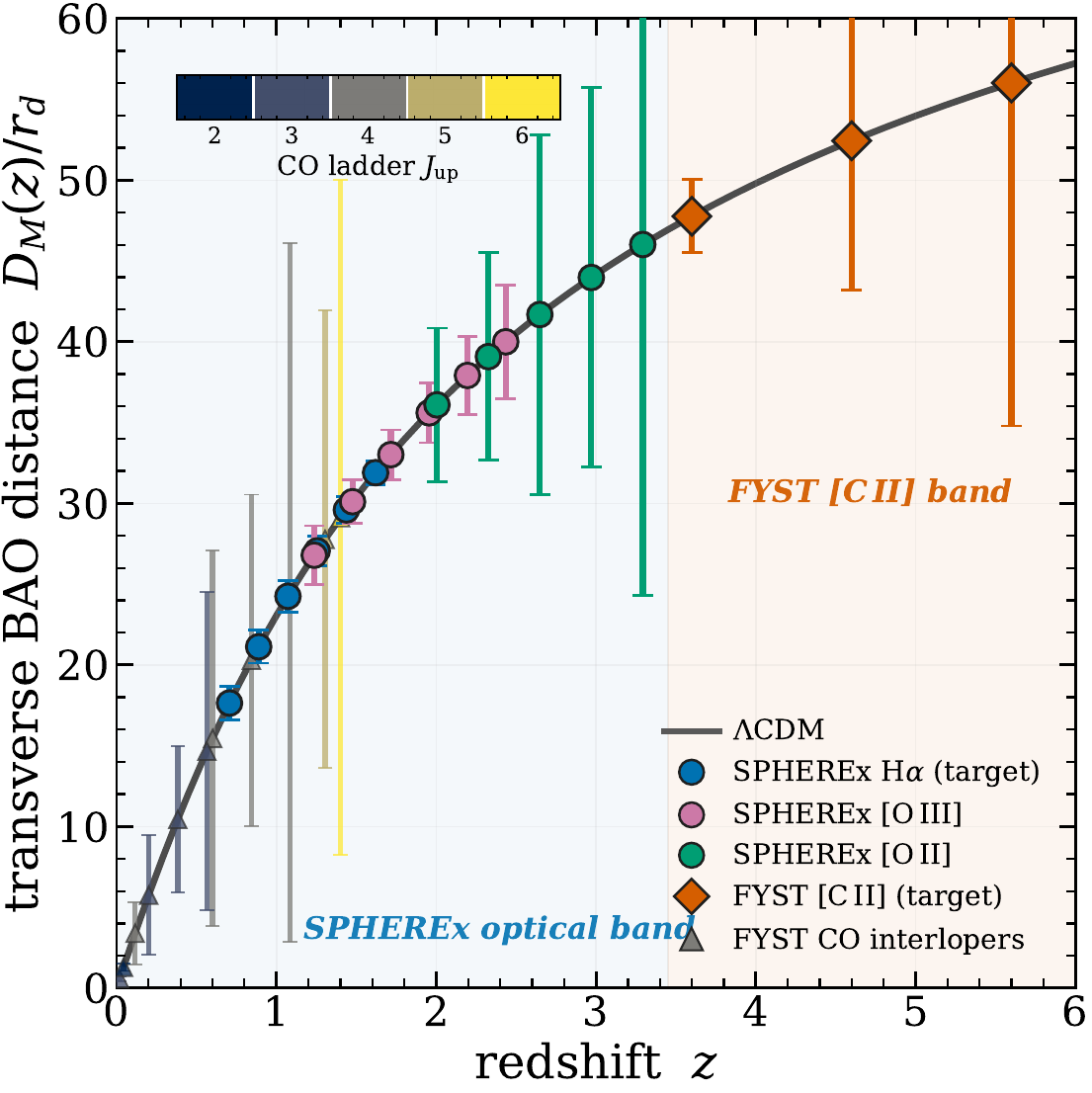}
  \caption{Transverse BAO distance $D_M(z)/r_d$ recovered by jointly modeling the target lines and their interlopers in two LIM bands. The optical band of SPHEREx provides \halpha{} target points and
\oiii{}, \oii{} interloper anchors. The FYST band adds high-redshift \cii{}
target points and lower redshift CO interlopers. Error bars are calculated with $20\%$ priors on line intensity
and bias. Modeling interlopers turns emission from other redshifts into additional BAO distance anchors at the redshifts selected by the line rest wavelengths.}
  \label{fig:bao_multiplex}
  \label{fig:bao_budget}
\end{figure}

\emph{Interlopers as BAO anchors.---}
The same array of lines that creates interloper confusion also fixes the redshifts at
which BAO features can be measured by using a single map at a single observed frequency channel. If these lines are modeled rather than
removed, it is possible to measure distances at several epochs. This observable is comparatively robust to line-amplitude errors: in a two-line limit, an amplitude error mainly changes the relative weight of the projected
components, while the dilation is set by the anisotropic projection shape. In
the SPHEREx and FYST tests, a $20\%$ interloper calibration error produces
sub-$0.5\sigma$ shifts in $(\alpha_\perp,\alpha_\parallel)$.

In Figure~\ref{fig:bao_multiplex}, we represent the resulting transverse BAO ladder. The
SPHEREx optical band provides \halpha{}, \oiii{}, and \oii{} anchors over
$z\simeq0.7$--$3.3$, while a FYST \cii{} band extends the ladder to
$z\simeq5.6$. In the adopted model, the optical interlopers provide most of the
added leverage, whereas the CO ladder is fainter and contributes weaker
low-redshift anchors. We omit the constraints from H$\beta$ for clarity because its anchor is
subdominant. Thus, emission from interlopers usually
treated as contamination becomes a dataset for transverse BAO at multiple redshifts by using only two observed bands. We verified that these decisions are stable to numerical prescriptions and choices of survey specifications:
varying finite-difference steps, $(k,\mu)$ binning, $k_{\max}$, noise level, beam size,
and priors on astrophysical modeling shifts the contours in
the expected direction, but does not change which parameters favor modeling and
which are limited by bias. These results highlight the redshift leverage gained by modeling interlopers. The precision of each anchor is model dependent, but the redshift
locations are fixed by atomic or molecular rest wavelengths, making the ladder
geometry robust.

\emph{Conclusions.---}
In this work, we show interloper removal is not automatically the optimal
method for cosmological analysis using LIM observations. Once the possible emitting lines, their redshifts, and calibration priors are included in the model, interlopers are additional maps of large-scale structure at other epochs, projected into the coordinate system of the target line. Removing them can therefore discard both their own cosmological information
and the target modes lost during the cleaning process. The optimal treatment is a decision problem rather than a universal preprocessing rule. An interloper contribution should be kept only when it improves precision and
keeps the induced bias below a chosen tolerance. Furthermore, in the
forecasts presented here, the BAO dilation parameters provide the cleanest
geometric test and remain comparatively robust to astrophysical calibration errors. The parameter $\Omega_m h^2$ also benefits from interloper modeling in the adopted setup, whereas $\sigma_8$ and the dark-energy parameters are more sensitive to the line model and priors, and can be biased by a tight but miscentered calibration.                               
The BAO example gives a concrete geometric realization of this decision
rule: the gain comes from adding distance anchors at the interloper
redshifts and from preserving target modes that cleaning would otherwise
discard. The magnitude of this gain depends on the line brightness model,
the spectral resolution, the survey noise, and the cost of cleaning. Modeling the lines in
just two observed bands yields a transverse BAO ladder spanning
$0.7\lesssim z\lesssim5.6$, with anchors pinned to atomic and molecular
rest wavelengths, and the gains are largest exactly where the target is
faintest.                                      

The present forecasts use empirical and halo-model line prescriptions, and
real data analyses must incorporate foregrounds, masks, transfer functions, and
non-Gaussian covariance. We do not impose a fixed dust attenuation in the fiducial forecast as attenuation and calibration uncertainty are absorbed into the marginalized nuisance parameters \cite{Roy2026-dust}. These effects can move the decision boundary, but they do not change the central claim of this work. The decision criterion itself is
inexpensive because a Fisher-level computation can be run quickly for deciding the data analysis strategy. Since SPHEREx is already collecting data and FYST construction is underway, future LIM surveys will integrate interlopers into estimators by designing how and when to excise them from the dataset and when to calibrate them as cosmological signal.

\begin{acknowledgments}
\emph{Acknowledgments.---}AR acknowledges support from NASA under award No.\,80NSSC18K1014939.
AR thanks Nick Battaglia, Anthony Pullen, Rachel Somerville, and David Spergel
for useful discussions, and Liana Giglio for support and encouragement during
this work. AR also thanks Manush Manju for reviewing the draft.
The Flatiron Institute is supported by the Simons Foundation.
\end{acknowledgments}

\onecolumngrid
\section*{End Matter}
\twocolumngrid
\subsection*{Forecast and projection model}
\label{app:pipeline}
We work in the comoving frame of the target line and fix a target-redshift
range together with an observed frequency (or wavelength) channel.  Each
rest-frame line $L$ in the catalog then contributes from the redshift
\begin{equation}
1+z_L=\frac{\nu_{\rm rest,L}}{\nu_{\rm obs}}
        =\frac{\lambda_{\rm obs}}{\lambda_{\rm rest,L}} .
\label{eq:app_zline}
\end{equation}
Lines falling outside the survey redshift range or the instrument's spectral
coverage are dropped.  Since the surviving identities and redshifts follow from
the rest-frame wavelengths rather than from any statistical assignment, the
multi-redshift data model is fixed from the outset.
An interloper at redshift $z_L$ enters the target-frame map with a geometric
distortion set by
\begin{equation}
q_{\perp,L}=\frac{D_M(z_L)}{D_M(z_{\rm t})},\qquad
q_{\parallel,L}=\frac{y(z_L)}{y(z_{\rm t})},
\label{eq:app_q}
\end{equation}
where $D_M(z)$ is the transverse comoving distance and
\begin{equation}
y(z)=\frac{d\chi}{d\nu_{\rm obs}}
    =\frac{c(1+z)^2}{H(z)\nu_{\rm rest}} .
\label{eq:app_y}
\end{equation}
Thus the wavevector entering the physical line power is
\begin{equation}
k_{\perp,L}=\frac{k_\perp}{q_{\perp,L}},\qquad
k_{\parallel,L}=\frac{k_\parallel}{q_{\parallel,L}},\qquad
k_L=\sqrt{k_{\perp,L}^2+k_{\parallel,L}^2},
\label{eq:app_kmap}
\end{equation}
with direction cosine $\mu_L=k_{\parallel,L}/k_L$.
The prefactor $(q_{\perp,L}^2q_{\parallel,L})^{-1}$ in
Eq.~(\ref{eq:pobs}) is the Jacobian from the interloper's true volume element to
the target-frame volume.  Each line is thus evaluated at its own redshift and
its own distorted $(k,\mu)$ before the contributions are summed; the observed
power is a projected sum, not a same-redshift sum of powers.
For each line we use the large-scale redshift-space intensity power
\begin{equation}
P_L(k,\mu,z)=I_L^2
\left[b_L(z)+f(z)\mu^2\right]^2P_{\rm m}(k,z)
+P_{{\rm shot},L},
\label{eq:app_line_power}
\end{equation}
with mean intensity $I_L$, luminosity-weighted bias $b_L$,
and shot noise $P_{{\rm shot},L}$.  The flagship SPHEREx run takes
$P_{\rm m}(k,z)$ from the CLASS code and forms finite differences in
$(\Omega_{\rm m},\sigma_8,h,w_0, w_a)$.  Every line additionally carries nuisance
parameters $(\ln I_L,\ln b_L,\ln P_{{\rm shot},L})$, over which the
quoted cosmological constraints are marginalized.  The amplitudes $I_L$,
$b_L$, and $P_{{\rm shot},L}$ are not free parameters but moments of a
halo-luminosity model; the star-formation law, the optical and sub-mm line
scalings, and the forecast pipeline are summarized in this End Matter.

\subsection*{Line model}
\label{app:line_model}
The line amplitudes that we use in Eq.~(\ref{eq:pobs}) are generated from a common
halo-SFR model spanned across the redshifts of interest.  For a
line label $L$, such as ${H\alpha}$,  \cii{}, etc., we use the moments of halo luminosity $\mathcal{L}_L(M,z)$, halo mass function
$dn/dM$, and halo bias $b_h(M,z)$, as
\begin{align}
\rho_{L}(z)&=\int dM\,\frac{dn}{dM}(M,z)\mathcal{L}_L(M,z),\\
b_L(z)&=\rho_L^{-1}\int dM\,\frac{dn}{dM}(M,z)
\mathcal{L}_L(M,z)b_h(M,z),\\
P_{{\rm shot},L}(z)&={\cal C}_L^2(z)
\int dM\,\frac{dn}{dM}(M,z)\mathcal{L}_L^2(M,z)\,S_{\rm sc},
\label{eq:app_line_moments}
\end{align}
with $I_L(z)={\cal C}_L(z)\rho_L(z)$.  Here, ${\cal C}_L$ converts luminosity to
intensity, and the term $S_{\rm sc}$ accounts for lognormal luminosity scatter.  We use a
Sheth--Tormen mass function and bias to calculate these quantities \cite{ShethTormen1999}. The SFR in different halo masses across a redshift range
is represented by a double power-law, as prescribed by UniverseMachine
\cite{Behroozi2013AverageSFH,Behroozi2019UniverseMachine}, which can be written as
\begin{equation}
{\rm SFR}(M,z)=
\epsilon(z)\left(\frac{M}{M_p(z)}\right)^{1.3}
\left[1+\left(\frac{M}{M_p(z)}\right)^{1.6}\right]^{-1},
\label{eq:app_sfr}
\end{equation}
where $M_p(z)$ can be represented as $3\times10^{11}[(1+z)/3]^{-0.4}\,M_\odot$ and
$\epsilon(z)=2[(1+z)/3]^{1.2}\,M_\odot\,{\rm yr}^{-1}$.  

The optical luminosity scalings follow the intrinsic SFR-luminosity conversions collected by Gong et al.~\cite{Gong2016OpticalLines},
\begin{align}
\mathcal{L}_{H\alpha}&=3.31\times10^7\,{\rm SFR},\\
\mathcal{L}_{H\beta}&=0.35\,\mathcal{L}_{H\alpha},\\
\mathcal{L}_{\rm [OIII]}&=3.44\times10^7\,{\rm SFR},\\
\mathcal{L}_{\rm [OII]}&=1.87\times10^7\,{\rm SFR},
\label{eq:app_optical_lum}
\end{align}

Here, the unit of luminosity is in solar luminosities. The \halpha{} normalization follows the Kennicutt SFR
calibration \cite{Kennicutt1998SFR}, \hbeta{} uses the Balmer ratio for case-B, and
the \oii{} and \oiii{} scalings follow optical LIM modeling as described in
\citet{Gong2016OpticalLines}.  For the sub-mm lines, we use empirical \cii{}--SFR and CO--IR relations.
For \cii{} line, we adopt the starburst calibration of \citet{DeLooze2014FIRSFR}
\begin{equation}
\log_{10}\!\left(\frac{{\rm SFR}}{M_\odot\,{\rm yr}^{-1}}\right)
=
-7.06+
\log_{10}\!\left(\frac{\mathcal{L}_{\rm [CII]}}{L_\odot}\right).
\label{eq:cii_lum}
\end{equation}
For CO lines, we connect SFR to infrared luminosity and then to CO brightness-temperature relations,
\begin{align}
{\rm SFR}
&=10^{-10}\frac{L_{\rm IR}}{L_\odot}\,
M_\odot\,{\rm yr}^{-1},\\
\log_{10}\!\left(\frac{L_{\rm IR}}{L_\odot}\right)
&=\alpha_{\rm CO}
\log_{10}\!\left(
\frac{L'_{\rm CO(1-0)}}{{\rm K\,km\,s^{-1}\,pc^2}}
\right)+\beta_{\rm CO},\\
\mathcal{L}_{\rm CO(J,J-1)}
&=4.9\times10^{-5}\,
J^3 r_{J1}
\left(\frac{L'_{\rm CO(1-0)}}{{\rm K\,km\,s^{-1}\,pc^2}}\right)
L_\odot .
\label{eq:co_lum}
\end{align}
Here $\alpha_{\rm CO}=1.37$, $\beta_{\rm CO}=-1.74$, and
$r_{J1}=L'_{\rm CO(J,J-1)}/L'_{\rm CO(1-0)}$ is the adopted CO SLED ratio.
These relations follow \citet{DeLooze2014FIRSFR} for \cii{}, \citet{Li2016COIM} for the IR--CO
conversion, and \citet{Mashian2015CO} for the multi-$J$ CO ladder
\cite{DeLooze2014FIRSFR,Li2016COIM,Mashian2015CO}.

\subsection*{Experiment specifications}
\label{app:experiments}
Instrumental sensitivity enters the forecast through a single number, the
white noise power spectrum, $P_N=\sigma_{\rm vox}^{2}V_{\rm vox}$, where $V_{\rm vox}$ is the
comoving voxel volume determined by the beam and the spectral channel size.
SPHEREx is a wide field spectrophotometer whose $6.2''$ pixels and $R=41$
spectral channels produce small voxels, $V_{\rm vox}=1.27\,{\rm Mpc^{3}}$ at
$z=1$, so its per pixel surface brightness depth converts directly into a noise
power spectrum: $16.5$ and $2.3\;{\rm nW\,m^{-2}\,sr^{-1}}$ for the all-sky
and deep surveys~\cite{Spherex-Bock-2025} give $P_N=3.4\times10^{2}$ and
$6.5\;({\rm nW\,m^{-2}\,sr^{-1}})^{2}\,{\rm Mpc^{3}}$ respectively.  We target
H$\alpha$ at $z=1$, observed at $1.313\,\mu$m, a
patch spanning $0.9<z<1.1$. On the other hand, FYST occupies the [CII] target line at higher redshift regime.  Its $48''$ beam and $280$~GHz channels make
voxels two orders of magnitude larger, $115\,{\rm Mpc^{3}}$ at $z=5.8$, and the
EoR-Spec specifications therefore quote the noise power itself rather than a
per-pixel depth: $N_{\rm white}=4.9\times10^{9}\,
{\rm Mpc^{3}\,Jy^{2}\,sr^{-2}}$~\cite{FYST} for the $280$~GHz channel, which
places \cii{} at $z=5.79$, equivalent to $2.7\,\mu$K per voxel.

\bibliographystyle{apsrev4-2}
\bibliography{references}
\end{document}